\documentclass[aps,prd,preprint,groupedaddress,showpacs]{revtex4}

\usepackage{epsfig}
\usepackage{graphics}

\begin{document}

\leftmargin -2cm
\def\choosen{\atopwithdelims..}
~~\\
DESY 06-231\\
December 2006\\
hep-ph/0612***

\boldmath
\title{On relative contributions of fusion and fragmentation mechanisms \\
in $J/\psi$ photoproduction at high energy}
\unboldmath

   \author{\firstname{V.A.} \surname{Saleev}}
\email{saleev@mail.desy.de, saleev@ssu.samara.ru}
   \thanks{on leave from Department of Physics, Samara
 State University, Ac. Pavlov St. 1,
443011 Samara, Russia}
  \affiliation{{II.} Institut f\"ur Theoretische Physik, Universit\" at Hamburg,
Luruper Chaussee 149, 22761 Hamburg, Germany}

\author{\firstname{A.V.} \surname{Shipilova}}
\email{alexshipilova@yandex.ru}
\affiliation{Department of Physics, Samara State University,
Ac.\ Pavlov St.~1, 443011 Samara, Russia}

\begin{abstract}
We study $J/\psi$ photoproduction via the fusion and fragmentation mechanisms at the HERA Collider
within the frameworks of the collinear parton model and the quasi-multi-Regge
kinematics approach using the factorization formalism of
non-relativistic QCD at leading order in the strong-coupling
constant $\alpha_s$ and the relative velocity $v$ of the bound
quarks.
It is shown that  the fusion production mechanism dominates
over the fragmentation production mechanism at the all relevant $J/\psi$ transverse momenta.
  The $J/\psi$ meson $p_T-$spectra in the fragmentation and fusion production
  at the asymptotilally large $p_T$ have equal slopes in the quasi-multi-Regge
kinematics approach, opposite the collinear parton model.

\end{abstract}


\maketitle \maketitle

\section{Introduction}

The phenomenology and the theory of processes involving the production
 of heavy quarkonia have been vigorously developed for last decade following the
 measurement of the transverse-momentum ($p_T$) spectra of prompt $J/\psi$ mesons
 by the CDF Collaboration at the Tevatron Collider \cite{CDFcharm} and
 by the ZEUS and H1 Collaboration at the HERA Collider \cite{HERAH1,HERAZEUS}.
 The color-singlet model \cite{CSM} previosly proposed to describe a non-perturbative
  transition  of a $Q\bar Q$ pair to final-state quarkonium was extended in a
  natural way within the formalism of nonrelativistic QCD (NRQCD) \cite{NRQCD}.
  The extended model takes into account $Q\bar Q$-pair production not only in the
  color-singlet state but also in the color-octet state.

  The NRQCD formalism makes it possible to calculate consistently, by perturbation
   theory in two small parameters (strong coupling constant $\alpha_s$ at the
   scale of the heavy-quark mass and the relative velocity $v$ of quarks in quarkonium),
   not only the parton cross sections for quarkonium production processes through the fusion
   of $Q$ and $\bar Q$ quarks but also the universal fragmentation functions for parton
   splitting into various quarkonium states \cite{FFNRQCD}.

   Usually it is assumed that, in the region of large quarkonium transverse momenta
   ($p_T\gg M$, where $M$ is the quarkonium mass), the fragmentation production mechanism
    is more adequate, because it takes into account effectively high order
    corrections in $\alpha_s$, than the mechanism associated with the fusion of a heavy
    quark and a heavy antiquark produced in a hard subprocess, which is calculated
    in the lowest order (LO) in  strong coupling constant $\alpha_s$. However, the numerical
    value of the critical transverse momentum is unknown a'priori.

In the energy region of the Tevatron and HERA Colliders the main
contribution to the heavy quarkonium production cross sections
comes from the gluon-gluon or the photon-gluon fusion at the small
values of the argument $x$ of the gluon distribution function. In
the conventional collinear parton model \cite{PartonModel}, the
initial-state gluon dynamics is controlled by the
Dokshitzer-Gribov-Lipatov-Altarelli-Parisi (DGLAP) evolution
equation \cite{DGLAP}. In this approach, it is assumed that $S >
\mu^2 \gg \Lambda_{\rm QCD}^2$, where $\sqrt{S}$ is the invariant
collision energy, $\mu$ is the typical energy scale of the hard
interaction, and $\Lambda_{\rm QCD}$ is the asymptotic scale
parameter of QCD. In this way, the DGLAP evolution equation takes
into account only one big logarithm, namely $\ln(\mu/\Lambda_{\rm
QCD})$. In fact, the collinear-parton approximation is used, and
the transverse momenta of the incoming gluons are neglected.

In the high-energy limit, the contribution from the partonic subprocesses
involving $t$-channel gluon exchanges to the total cross section can become
dominant.
The summation of the large logarithms $\ln(\sqrt{S}/\mu)$ in the evolution
equation can then be more important than the one of the
$\ln(\mu/\Lambda_{\rm QCD})$ terms.
In this case, the non-collinear gluon dynamics is described by the
Balitsky-Fadin-Kuraev-Lipatov (BFKL) evolution equation \cite{BFKL}.
In the region under consideration, the transverse momenta ($k_T$) of the
incoming gluons and their off-shell properties can no longer be neglected,
and we deal with reggeized $t$-channel gluons.
The theoretical frameworks for this kind of high-energy phenomenology are the
$k_T$-factorization approach \cite{KTGribov,KTCollins} and the
(quasi)-multi-Regge kinematics (QMRK) approach
\cite{KTLipatovFadin,KTAntonov}. Last one is based on effective quantum field
theory implemented with the non-abelian gauge-invariant action, as suggested
a few years ago \cite{KTLipatov}.
However, the $k_T$-factorization approach has well-known principal
difficulties \cite{smallx} at next-to-leading order (NLO).
By contrast, the QMRK approach offers a conceptual solution of the NLO
problems \cite{Ostrovsky}.

Our previous analysis of $J/\psi$ meson production at the Fermilab
Tevatron Colliders using the high-energy factorization scheme
and the NRQCD formalism\cite{KniehlSaleevVasin1,PRD2003,SaleevVasin} has
shown the dominant role of the color-octet intermediate state $^3S_1^{(8)}$
 contribution at the large $p_T$ region in the
fusion production and in the fragmentation production with the gluon splitting.
We found also the good agreement between the values of non-perturbative matrix
element (NME) $\langle {\cal O}^{J/\psi}[^3S_1^{(8)}]\rangle$ obtained by the fit
in collinear parton model and in the
QMRK approach. The similar analysis for the $J/\psi$ photoproduction at the high energy
is the subject of study in this paper. The situation should be different because of the $c-$quark
fragmentation via the color-singlet state dominates over the gluon fragmentation via the color-octet state
in $J/\psi$ photoproduction.

This paper is organized as follows.
In Sec.~\ref{sec:two} we discuss basic ideas and formulas of the fragmentation approach,
including the Dokshitzer-Gribov-Lipatov-Altarelli-Parisi (DGLAP) evolution equations
for the $c-$quark and gluon fragmentation functions into $J/\psi$ meson.
In Sec.~\ref{sec:three} we present relevant LO squared amplitude in the QMRK approach
 and master formulas for the differential cross sections.
In Sec.~\ref{sec:four} we discuss our results for the $J/\psi$ photoproduction at the HERA Collider.
In Sec.~\ref{sec:five} we summarize our conclusions.

\boldmath
\section{Fragmentation model}
\unboldmath
\label{sec:two}

Within the framework of the NRQCD, the fragmentation function for quarkonium $H$ production
can be expressed as a sum of terms, which are factorized into a short-distance coefficient
and a long distant matrix element \cite{NRQCD}:
\begin{equation}
D(a\to H)=\sum_n D(a\to Q\bar Q[n])\langle {\cal O}^{H}[n]\rangle.
\end{equation}
Here the n denotes the set of color and angular momentum numbers of the $Q\bar Q$ pair, in which
 the fragmentation function is $D(a\to Q\bar Q[n])$. The last one can be calculated perturbatively
 in the strong coupling constant $\alpha_s$.
The non-perturbative transition of the $Q\bar Q$ pair into ${\cal H}$ is
described by the NMEs $\langle {\cal O}^{\cal H}[n]\rangle$, which can be
extracted from experimental data.
To LO in $v$, we need to include the $c\bar c$ Fock states
$n = {^3S}_1^{(1)}, {^3S}_1^{(8)}, {^1S}_0^{(8)}, {^3P}_J^{(8)}$ if
${\cal H} = J/\psi (\psi^\prime)$.
The recipes of calculations for heavy quarkonium fragmentation functions
at the initial scale $\mu=\mu_0=2m_c$ in LO NRQCD formalism are well known \cite{FFNRQCD}
and detailed formulas for the  gluon and $c-$quark fragmentation via the
different intermediate states are presented,
for example, in Ref.\cite{KniehlKramer}.

\begin{figure}[ht]
\begin{center}
\includegraphics[width=0.8\textwidth, clip=]{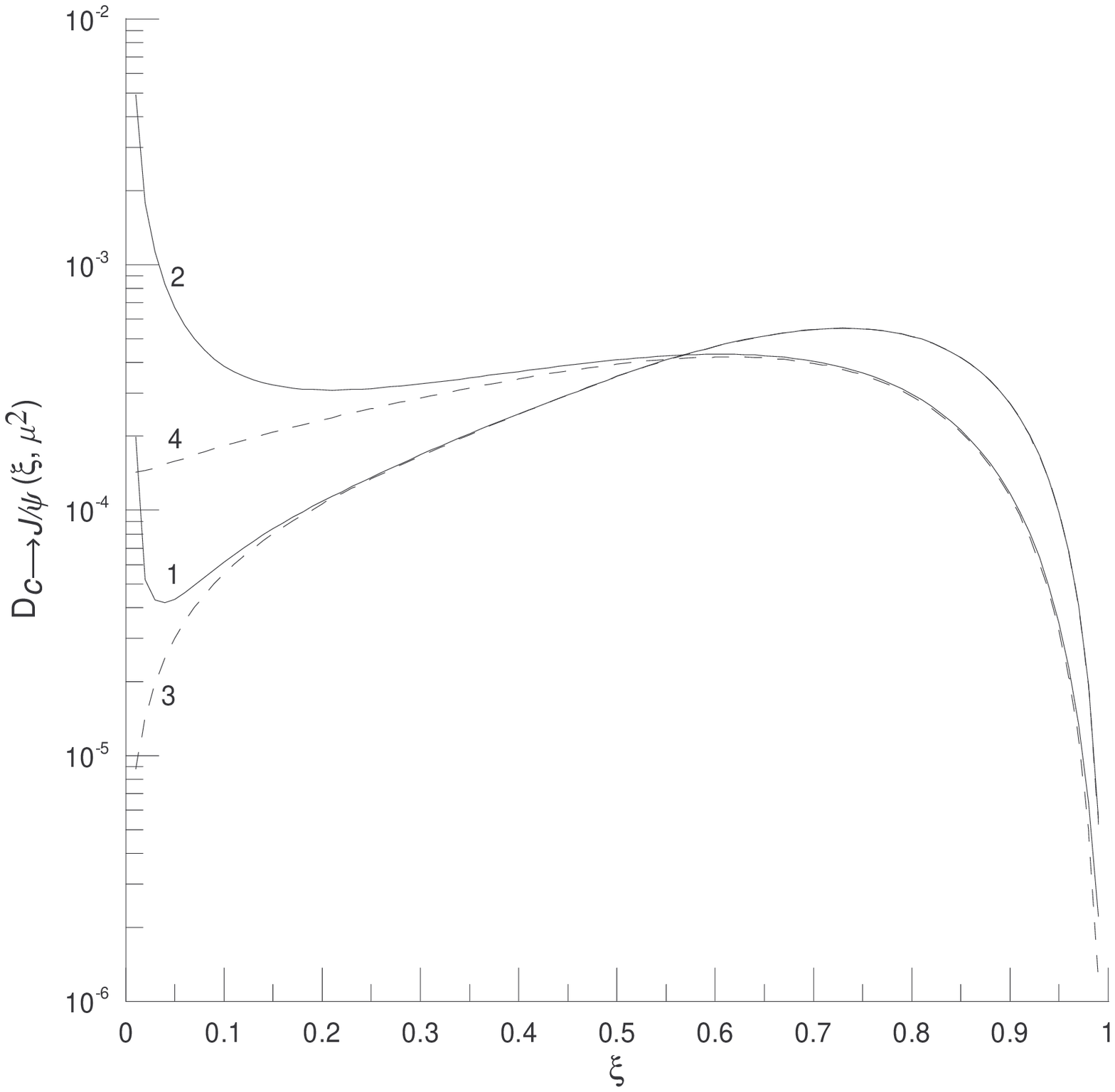}
\end{center}
\caption{The fragmentation function $D_{c\to J/\psi}(z,\mu^2)$ at the $\mu^2=10$
 GeV$^2$ (the curves 1 and 3) and $\mu^2=300$ GeV$^2$ (the curves 2 and 4).
The solid lines are results of Eqs.~(\ref{eq:evol}), the dashed are obtained
 neglecting the $c-g$ coupling. \label{fig:cfrag}}
\end{figure}

\begin{figure}
\begin{center}
\includegraphics[width=0.8\textwidth, clip=]{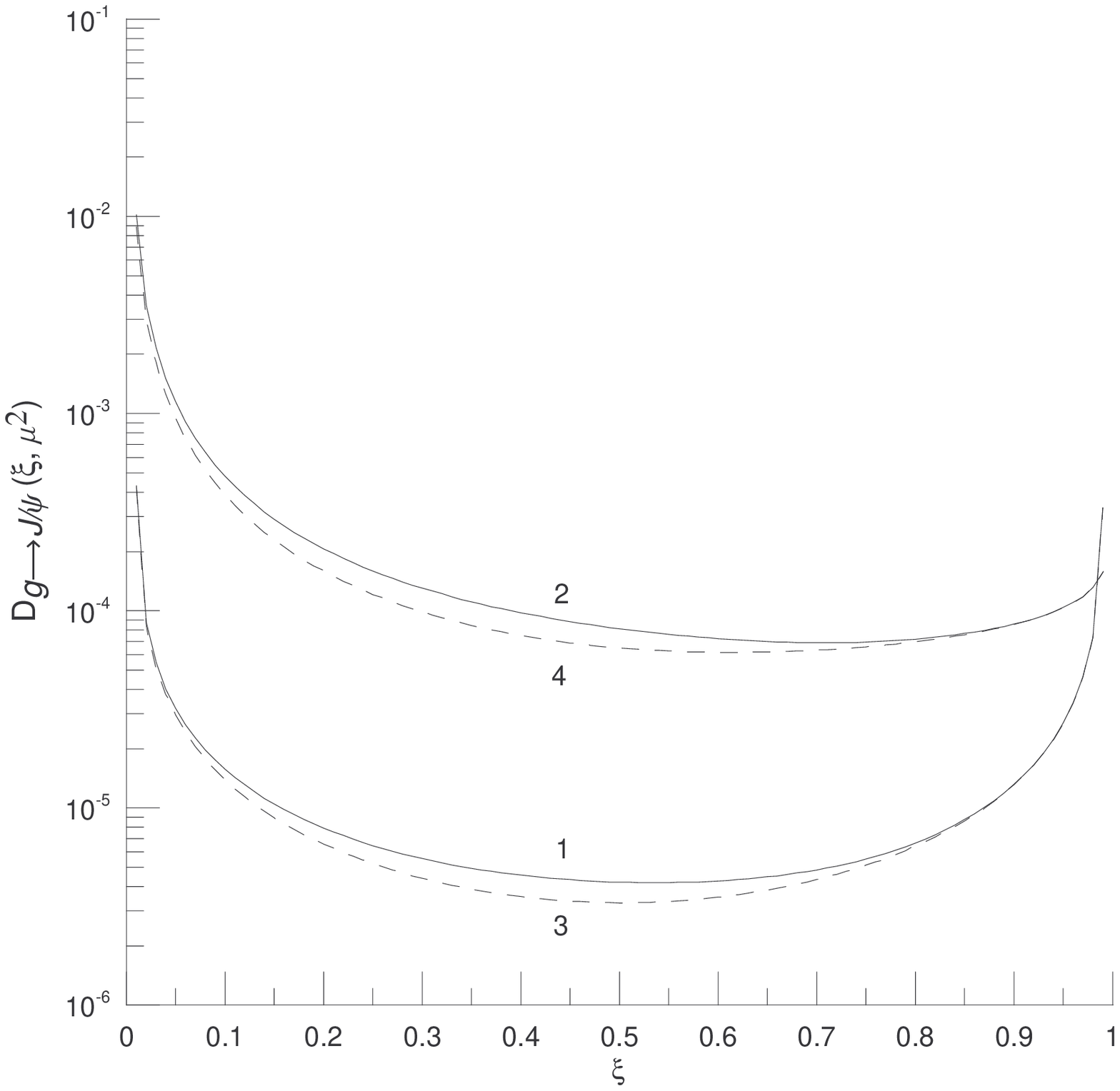}
\end{center}
\caption{The fragmentation function $D_{g\to J/\psi}(z,\mu^2)$ at the $\mu^2=10$
 GeV$^2$ (the curves 1 and 3) and $\mu^2=300$ GeV$^2$ (the curves 2 and 4).
The solid lines are results of Eqs.~(\ref{eq:evol}), the dashed are obtained
neglecting the $c-g$ coupling.\label{fig:gfrag}}
\end{figure}

To calculate the $J/\psi$ production spectra, we need the fragmentation functions
 at the factorization scale $\mu\gg M$. Then, large logarithms in $\mu/M$ appear,
 which have to be ressumed. It is achieved by using the DGLAP equations
 \begin{equation}
 \frac{\mu^2d}{d\mu^2}D_{a\to H}(\xi,\mu)=\sum_b \int_\xi^1\frac{dx}{x}P_{ba}(\frac{\xi}{x},\alpha_s(\mu))
 D_{b\to H}(x,\mu),\label{eq:evol}
 \end{equation}
 where $\displaystyle{P_{ba}(\frac{\xi}{x},\alpha_s(\mu))}$  are the timelike splitting functions
 of parton $a$ into parton $b$. The initial conditions $D_{a\to H}(\xi,\mu_0)$ for $a=c,g$ and $H=J/\psi(\psi^\prime)$
 are taken from Ref.~\cite{KniehlKramer}. The initial light-quark fragmentation functions are set equal to zero.
 They are generated at larger scale $\mu$. However, in the case of $J/\psi$ production at the HERA Collider
   their  contribution is very small. Here, $a=g,c,\bar c$. To illustrate the $\xi$ and $\mu$ dependence of the
   relevant
   fragmentation functions we show in Figs.~\ref{fig:cfrag} and \ref{fig:gfrag}
    the $c-$quark and the gluon fragmentation functions at the different
   values of $\mu$. We use LO in $\alpha_s$  approximation for the splitting functions. The dashed curves
 are obtained neglecting the $c-g$ coupling during the evolution ($a=c$ or $a=g$),
 the solid curves are obtained using the system
 (\ref{eq:evol}), which takes into account cross transitions between gluons and $c-$quarks during the evolution.

\boldmath
\section{Basic formulas}
\label{sec:three}
\unboldmath

Because we take into account the gluon and $c-$quark fragmentation into the $J/\psi$ mesons,
we need to sum contributions from LO partonic subprocesses in which these ones are produced.

In the collinear parton model we take into consideration the following subprocesses:
\begin{eqnarray}
\gamma + g &\to & c + \bar c,\label{eq:PMsub1}\\
\gamma + q (\bar q)&\to & g +q (\bar q), \label{eq:PMsub2}
\end{eqnarray}
where $q=u,d,s$. The squared amplitudes of the subprocesses (\ref{eq:PMsub1}) and
(\ref{eq:PMsub2}) are well known.

\begin{figure}[ht]
\begin{center}
\includegraphics[width=0.9\textwidth, clip=]{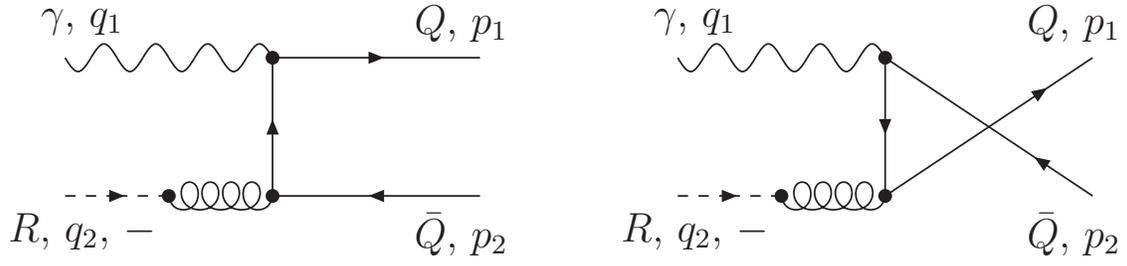}
\end{center}
\caption{The Feynman diagrams of the process $\gamma+R\to c+\bar
c$ }
 \label{fig:yRQQ}
\end{figure}

In the case of QMRK approach we have only one dominant LO partonic subprocess:

\begin{equation}
 \gamma + R \to c + \bar c, \label{eq:QMRK}
\end{equation}
which is described by the Feynman diagrams shown in Fig.~\ref{fig:yRQQ}.
Here $R$ is the reggeized gluon. Accordingly \cite{KTAntonov},
amplitude of the subprocess (\ref{eq:QMRK}) can be presented as follows:
\begin{eqnarray}
{\cal A}(\gamma+R\to c+\bar c)&=&-ig_se_c e T^a\bar U(p_1,m_c)[
\gamma^\alpha\frac{\hat p_1-\hat q_1+m_c}{(p_1-q_1)^2-m_c^2}\gamma^\beta+\nonumber\\
 &&+\gamma^\beta\frac{\hat p_1-\hat q_2+m_c}{(p_1-q_2)^2-m_c^2}\gamma^\alpha\label{eq:ampQMRK}
]V(p_2,m_c)\varepsilon^\alpha(q_1)(n^-)^\beta,
\end{eqnarray}
where $(n^-)^\mu=P_N^\mu/E_N$, $P_N^\mu=E_N(1,0,0,1)$ is the four-momentum of the proton,
$E_N$ is the energy of the proton, $\varepsilon^\alpha(q_1)$ is the polarization four-vector of the photon.
For any four-momentum $k^\mu$, we define $k^-=(k\cdot n^-)$.
Note, that reggeized gluon four-momentum can be written as
 \begin{equation}
  q_2^\mu=x_2 E_N(n^-)^\mu+q_{2T}^\mu,
 \end{equation}
where $q_{2T}^\mu=(0,{\bf q}_{2T},0)$, $q_2^2=q_{2T}^2=-{\bf q}_{2T}^2$
 and $x_2$ is the fraction of the proton longitudinal momentum
passed on to the reggeized gluons.

The squared amplitude obtained from (\ref{eq:ampQMRK}) reads:
\begin{equation}
\overline{|{\cal A}(\gamma+R\to c+\bar c)|^2}=\frac{64\pi^2e_c^2\alpha_s\alpha}{\tilde t^2\tilde u^2}
[2m_c^2(\tilde tp_1^- -\tilde up_2^-)^2+{\bf q}_{2T}^2\tilde u\tilde t((p_1^-)^2+(p_2^-)^2)],\label{eq:amp2}
\end{equation}
where the bar indicates average (summation) over initial-state (final-state) spins and colors,
$\tilde u=\hat u-m_c^2$,
$\tilde t=\hat t-m_c^2$,
 $\hat u=(q_1-p_2)^2$, $\hat t=(q_1-p_1)^2$, $\alpha=e^2/4\pi$, $\alpha_s=g_s^2/4\pi$, $e_c=+2/3$.
The obtained result (\ref{eq:amp2}) satisfies gauge invariant conditions,
$$\lim_{{\bf q}_{2T}^2\to 0}|{\cal A}(\gamma+R\to c+\bar c)|^2=0,$$
 and agrees with the $k_T$-factorization formula from Ref.~\cite{Catani}
 if we take into account relation between amplitudes in the QMRK and the $k_T-$factorization approach
 \cite{KniehlSaleevVasin2}:
\begin{equation}
|{\cal A}_{KT}(\gamma+R\to c+\bar c)|^2=\frac{(x_2E_N)^2}{{\bf q}_{2T}^2}  |{\cal A}(\gamma+R\to c+\bar c)|^2.\nonumber
\end{equation}

At HERA, the cross section of prompt $J/\psi$ production was
measured in a wide range of the kinematic variables
$W^2=(P_N+q_1)^2$, $Q^2=-q_1^2$, $x_1=(P_N\cdot q_1)/(P_N\cdot k)$,
$z=(P_N\cdot p)/(P_N\cdot q_1)$, and $p_T$, where $P_N^\mu$, $k^\mu$,
$k^{\prime\mu}$, $q_1^\mu=k^\mu-k^{\prime\mu}$, and $p^\mu$ are the
four-momenta of the incoming proton, incoming lepton, scattered
lepton, virtual photon, and produced $J/\psi$ meson,
respectively, both in photoproduction \cite{HERAZEUS}, at small
values of $Q^2$, and deep-inelastic scattering (DIS) \cite{HERAH1},
at large values of $Q^2$. At sufficiently large values of $Q^2$,
the virtual photon behaves like a point-like object, while, at
low values of $Q^2$, it can either act as a point-like object
(direct photoproduction) or interact via its quark and gluon
content (resolved photoproduction). Resolved photoproduction is
only important at low values of $z$ and $p_T$. The subject of our study
is  large $p_T$ photoproduction processes at the $Q^2\approx 0$ only.

In the region of $p_T\gg m_c$ we consider $c$-quark and $J/\psi$-meson
as massless particles, so that the fragmentation parameter $\xi$ is related to
 the $c-$quark and
$J/\psi-$meson four-momenta as follows:
\begin{equation}
p^\mu=\xi p_{1}^\mu.
\end{equation}
We only used non-zero $c-$quark mass in the initial factorization scale
$\mu_0=M=2m_c$ and, correspondingly, in the definition of the $\mu=\sqrt{p_T^2+4 m_c^2}$.

Let us first present the relevant formula for the double differential
cross sections of $J/\psi$ direct photoproduction in the collinear parton model
via the partonic subprocess (\ref{eq:PMsub1}):
\begin{eqnarray}
\frac{d\sigma^{ep}}{dzdp_T^2}= \int^1_{x_{1,min}} dx_1 f_{\gamma/e}(x_1)
\int^1_{\xi_{min}} \frac{d\xi}{\xi}D_{c\to J/\psi}(\xi,\mu)
\frac{\overline{|{\cal M}(\gamma+g\to c+\bar c)|^2}x_2G(x_2,\mu)}{16\pi z(\xi-z)(x_1x_2S)^2},\label{eq:crossPM}
\end{eqnarray}
where
\begin{eqnarray}
 x_2=\frac{p_T^2}{zx_1S(\xi-z)},\quad x_{1,min}=\frac{p_T^2}{zS(1-z)},
  \quad \xi_{min}=z+\frac{p_T^2}{x_1zS}\nonumber,
 \end{eqnarray}
$G(x_2,\mu)$ is the gluon collinear distribution function in the proton, for which we use numerical CTEQ
parameterization \cite{CTEQ} and $f_{\gamma/e}(x_1)$ is the quasi-real photon flux.
 In the Weiz\"acker-Williams approximation, the latter takes the form
\begin{equation}
f_{\gamma/e}(x_1)=\frac{\alpha}{2\pi}\left[\frac{1+(1-x_1)^2}{x_1}\ln
\frac{Q_{\rm max}^2}{Q_{\rm min}^2}
+2m_e^2x_1\left(\frac{1}{Q_{\rm min}^2}-\frac{1}{Q_{\rm max}^2}\right)\right],
\end{equation}
where $Q_{\rm min}^2=m_e^2x_1^2/(1-x_1)$ and $Q^2_{\rm max}$ is determined by
the experimental set-up, {\it e.g.}\
$Q^2_{\rm max}=1$~GeV$^2$ \cite{HERAZEUS}.

 The formulae in the case of the partonic subprocess (\ref{eq:PMsub2})
looks like (\ref{eq:crossPM}). The $J/\psi$ mesons are produced also via decays of excited charmonium states
and $B-$mesons. In fact, only the contribution of the decays $\psi^\prime\to J/\psi X$ is large at the relevant
kinematic condition. In the approximation used here we can estimate this one simply:
\begin{equation}
d\sigma (ep\to J/\psi X)=d\sigma^{direct}(ep\to J/\psi X)+   d\sigma^{direct}(ep\to \psi^\prime X)
\mbox{Br}(\psi^\prime\to J/\psi X),
\end{equation}
where  $\mbox{Br}(\psi^\prime\to J/\psi X)$ is the inclusive decay branching fractions,
$d\sigma^{direct}(ep\to \psi^\prime X)$ can be obtained formally from  $d\sigma^{direct}(ep\to J/\psi X)$
after replacement $\langle {\cal O}^{J/\psi}[n]\rangle \to \langle {\cal O}^{\psi^\prime}[n]\rangle$.

In the QMRK approach the double differential cross sections of $J/\psi$ direct photoproduction
reads:
\begin{eqnarray}
   \frac{d\sigma^{ep}}{dzdp_T^2}&=&\frac{1}{8(2\pi)^2}\int d \phi_2\int \frac{d{\bf q}_{2T}^2}{{\bf q}_{2T}^2}
   \int dx_1 f_{\gamma/e}(x_1)\int \frac{d\xi}{\xi}
   D_{c\to J/\psi}(\xi,\mu)\times\nonumber\\
   &&\times\frac{(x_2E_N)^2\Phi(x_2,{\bf q}_{2T}^2,\mu)}
   {z(\xi-z)(x_1x_2S)^2}\overline{|{\cal A}(\gamma+R\to c+\bar c)|^2},\label{eq:crossQMRK}
   \end{eqnarray}
where
$\phi_2$ is the angle between ${\bf q}_{2T}$ and the transverse momentum
${\bf p}_T$ of the $J/\psi$ meson, $\Phi(x_2,{\bf q}_{2T}^2,\mu)$ is the unintegrated reggeized gluon
distribution function. The collinear and the
unintegrated gluon distribution functions are formally related as:
\begin{eqnarray}
x_2 G(x_2,\mu)=\int^{\mu^2}_0d{\bf q}_{2T}^2 \Phi(x_2,{\bf q}_{2T}^2,\mu),
\end{eqnarray}
so that the normalizations of Eqs. (\ref{eq:crossQMRK}) and (\ref{eq:crossPM}) agree.
In  our numerical calculations, we use the unintegrated gluon distribution function
 by Kimber, Martin, and Ryskin (KMR) \cite{KMR}, which gives the best description
 in the framework of   QMRK approach \cite{KniehlSaleevVasin1,KniehlSaleevVasin2}
 of the heavy quarkonium production spectra measured at the Tevatron \cite{CDFcharm,Tevatron} .

To compare predictions of the fragmentation and the fusion mechanisms we need to calculate $J/\psi$
photoproduction cross section using the fusion model too. In the case of the LO collinear parton model
calculation we use squared amplitudes for the color-singlet contribution from Ref.~\cite{CSM} and for the
color-octet contribution from Ref.~\cite{Beneke}. Working in the QMRK approach with the fusion model
we use analytical results obtained in Refs.~\cite{KniehlSaleevVasin1,SaleevPSI}. All numerical values
 of parameters are the same as in the Ref.~ \cite{KniehlSaleevVasin1}.

\section{Results}
\label{sec:four}

We now present and discuss our numerical results.

\begin{figure}[ht]
\begin{center}
\includegraphics[width=0.9\textwidth, trim=0cm 5cm 0cm 3cm]{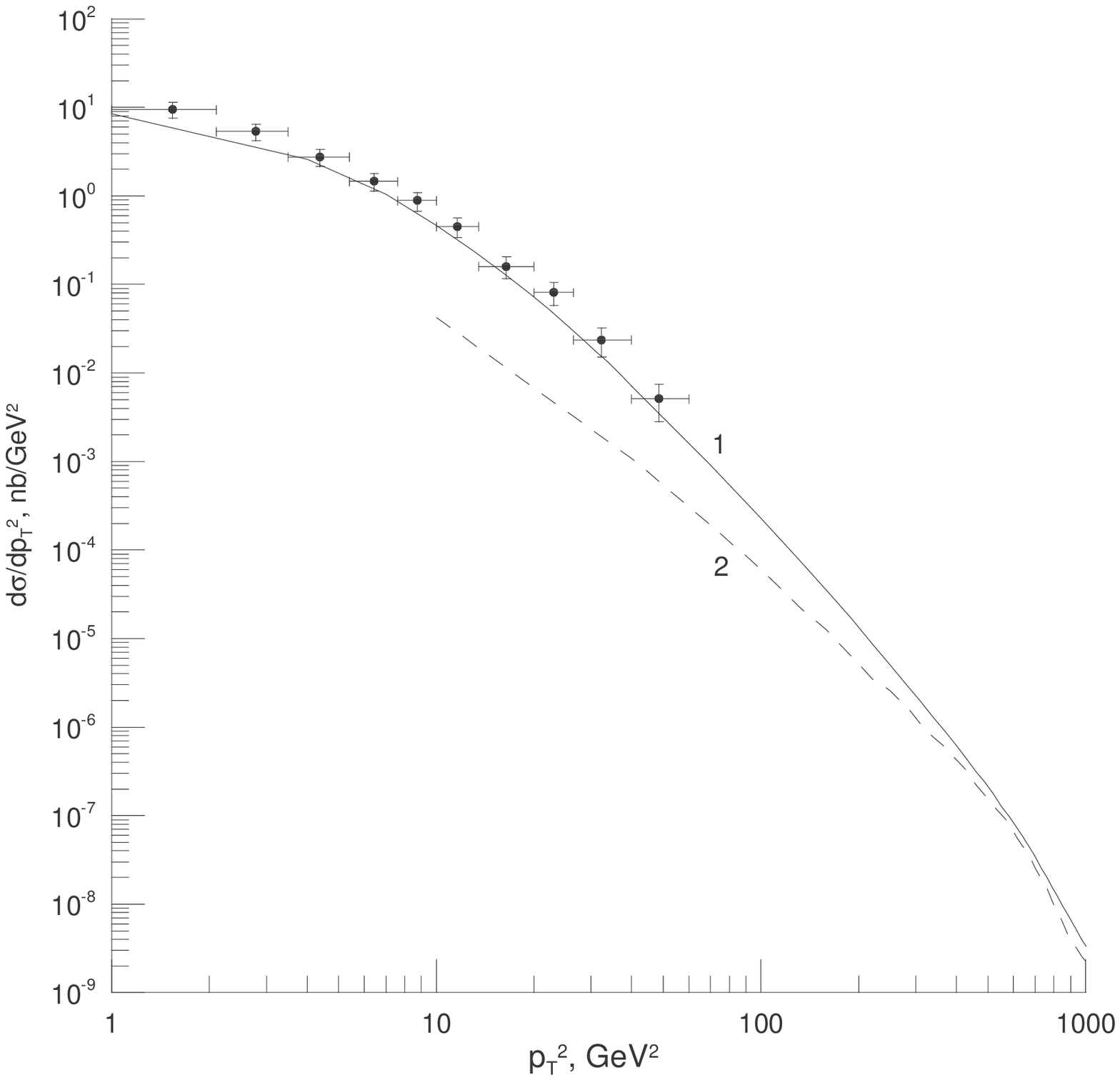}
\end{center}
\caption{$p_T$ spectrum of prompt $J/\psi$ in $ep$
scattering with $E_p=820$ GeV, $E_e=27.5$ GeV, 60 GeV $<W<$ 240 GeV, $Q^2<1$ GeV$^2$, and $0.3<z<0.9$.
The fusion production mechanism (curve 1), and the fragmentation production mechanism (curve 2) in the QMRK approach.
\label{fig:ptkt}}
\end{figure}

Let us compare $J/\psi$ meson $p_T$ spectra obtained using the fragmentation  and fusion models whithin
the QMRK approach. As it was shown recently \cite{SaleevPSI,KniehlSaleevVasin1}, we can describe  the
experimental data from HERA Collider \cite{HERAZEUS,HERAH1} in the fusion model without the color-octet contribution,
g.e. in the color-singlet model \cite{CSM} (see the curve 1 in the Fig.~\ref{fig:ptkt}).  The agreement
is good for the all measured $J/\psi$ transverse momenta up to $p_T^2=50$ GeV$^2$. The prediction of the fragmentation
model is shown in Fig. \ref{fig:ptkt} for $p_T^2> 10$ GeV$^2$ only (the curve 2). We see that the curve 2
tends to the curve 1 as $p_T^2$ increases but it is being lower at all relevant  $p_T^2$ till $p_T^2=10^3$ GeV$^2$.
The result is the same as in the hadronic $J/\psi$ production \cite{PRD2003,KniehlSaleevVasin1} although in the
photoproduction the $c-$quark fragmentation via the color-singlet state $^3S_1^{(1)}$ is dominant,
oppocite to the
hadroproduction where the gluon  fragmentation via the color-octet state $^3S_1^{(8)}$ is dominant.  So, we see
that in the QMRK approach the slope of the $p_T$ spectra sufficiently depends on the initial reggeized gluon
transverce momentum, which is ordered by the unintegrated distribution function,
 and both production mechanismes, the fusion and the fragmentation, have similar $p_T-$slopes
  at the large transverse momentum  of the $J/\psi$ meson.

\begin{figure}[ht]
\begin{center}
\includegraphics[width=0.9\textwidth, trim=0cm 5cm 0cm 3cm]{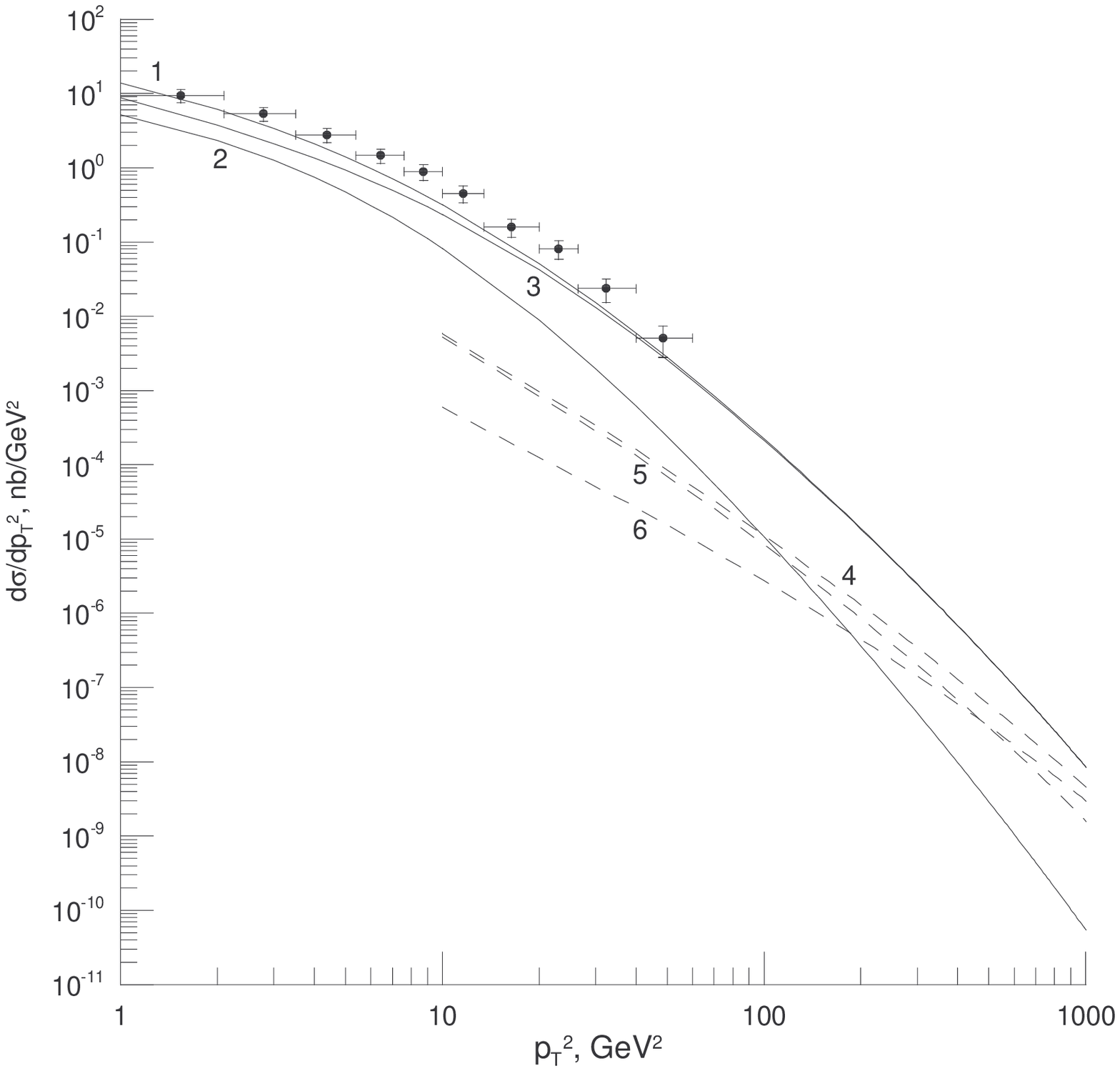}
\end{center}
\caption{$p_T$ spectrum of prompt $J/\psi$ in $ep$ scattering with
$E_p=820$ GeV, $E_e=27.5$ GeV, 60 GeV $<W<$ 240 GeV, $Q^2<1$
GeV$^2$, and $0.3<z<0.9$. The total contribution of the fusion
mechanism (curve 1), the color-singlet part of the fusion
mechanism (curve 2), the color-octet part of the fusion mechanism
(curve 3), the total contribution of the fragmentation mechanism
(curve 4), the color-singlet part of the fragmentation
 mechanism (curve 5), the color-octet part of the fragmentation mechanism (curve 6).
\label{fig:ptpm}}
\end{figure}

The situation in the relevant LO calculations within the collinear
parton model was studied earlier in
Refs.~\cite{KniehlKramer,Godbole}. It was shown that at
approximately $p_T^2 \ge 100$ GeV$^2$ the contribution of the
$c-$quark fragmentation production exeeds the contribution of the
fusion mechanism with the color-singlet LO amplitudes. Note, that
in both of the discussed approaches,
 the fragmentation and the fusion within the CSM in LO, do not describe data. However, if we
  take into account the
 color-octet contribution, with the values of the color-octet NMEs fixed at the Tevatron,
  we see that the fusion model describes data well, as it is shown
 in Fig.~\ref{fig:ptpm}
 (the curve 1). The contribution of the fragmentation model increases too, because of the additional
  piece coming from
 the gluon fragmentation via $^3S_1^{(8)}$ state in the partonic subprocesses $\gamma+q\to g+q$
 (the curve 6 in Fig.~\ref{fig:ptpm}). Moreover, at
 $p_T^2> 500$ GeV$^2$ the gluon fragmentation dominates over the $c-$quark fragmentation (the curve 5),
 but their sum (the curve 4) is
  still less than the fusion model prediction in the NRQCD approach up to  $p_T^2=2\cdot 10^3$ GeV$^2$.

\begin{figure}[ht]
\begin{center}
\includegraphics[width=1.0\textwidth, trim=0cm 5cm 0cm 3cm]{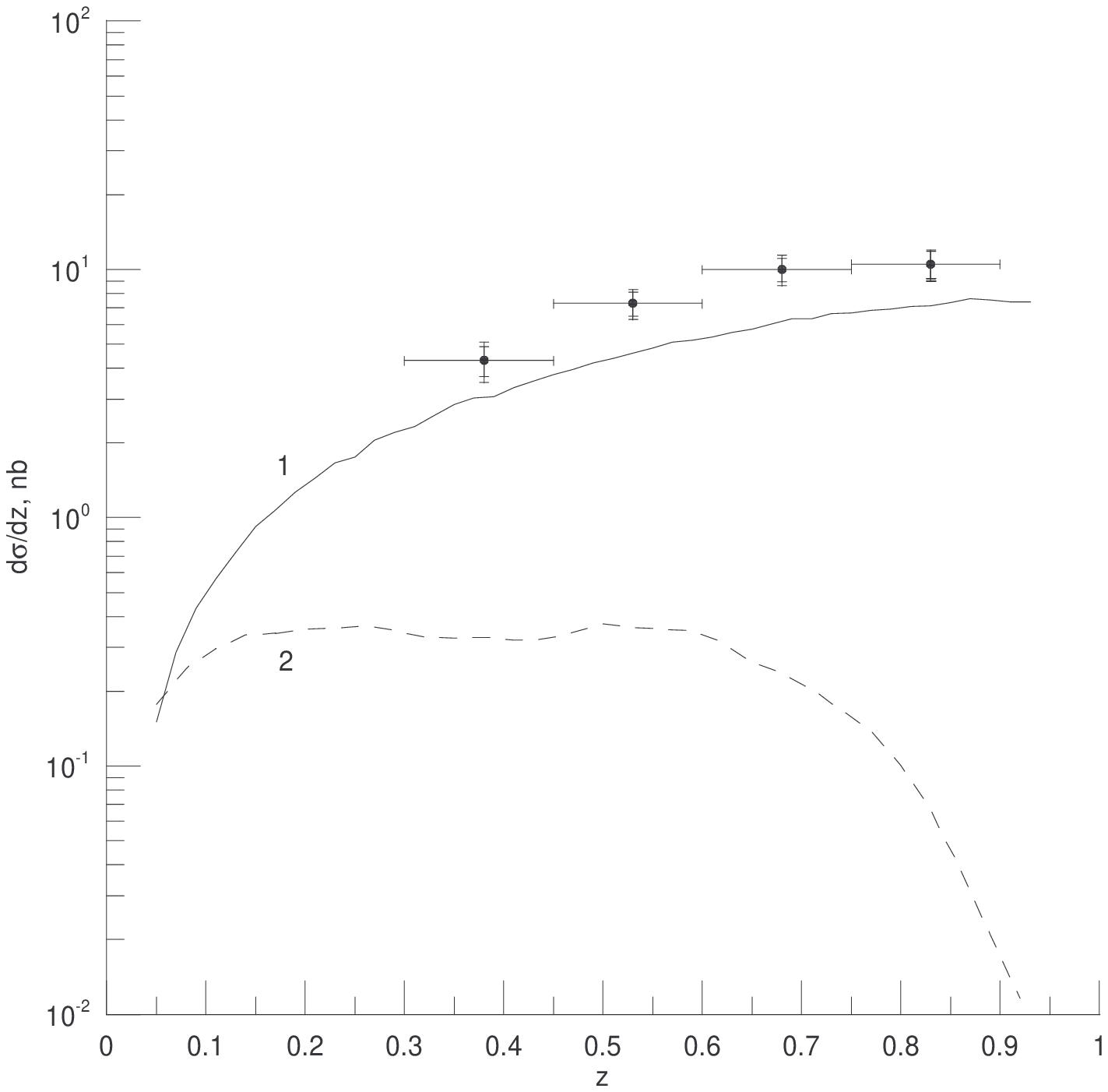}
\end{center}
\caption{ $z$ spectrum of prompt $J/\psi$ in $ep$
scattering with $E_p=820$ GeV, $E_e=27.5$ GeV, 60 GeV $<W<$ 240 GeV, $Q^2<1$ GeV$^2$, and $p_T>3$ GeV.
The fusion production mechanism (curve 1), and the fragmentation production mechanism (curve 2)
 in the QMRK approach\label{fig:zkt}}
\end{figure}

\begin{figure}[ht]
\begin{center}
\includegraphics[width=0.9\textwidth, trim=0cm 5cm 0cm 3cm]{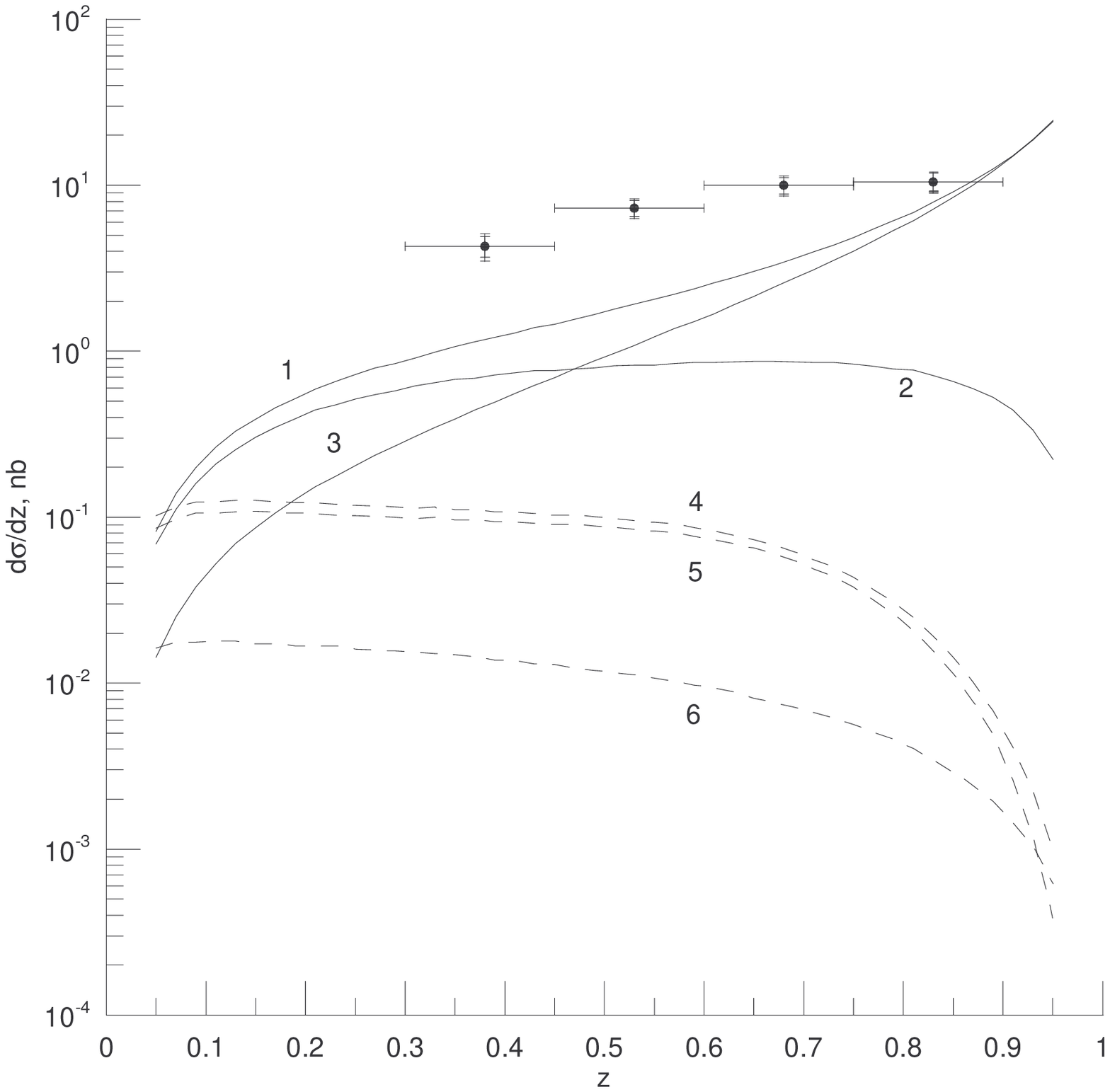}
\end{center}
\caption{ $z$ spectrum of prompt $J/\psi$ in $ep$ scattering with
$E_p=820$ GeV, $E_e=27.5$ GeV, 60 GeV $<W<$ 240 GeV, $Q^2<1$
GeV$^2$, and $p_T>3$ GeV. The total contribution of the fusion
mechanism (curve 1), the color-singlet part of the fusion
mechanism (curve 2), the color-octet part of the fusion mechanism
(curve 3), the total contribution of the fragmentation mechanism
(curve 4), the color-singlet part of the fragmentation
 mechanism (curve 5), the color-octet part of the fragmentation mechanism (curve 6).
\label{fig:zpm}}
\end{figure}

For completeness, we present here the $z-$spectra of $J/\psi$ calculated in the QMRK approach (Fig.~\ref{fig:zkt})
and in the collinear parton model   (Fig.~\ref{fig:zpm}) with the experimental data from HERA Collider
at the $p_T>3$ GeV \cite{HERAH1}. We see that in both approaches the fusion mechanism is dominant for all $z$,
 excepting the region of very small $z<0.1$. It is well known that at small $z$ the resolved $J/\psi$ meson
 photoproduction  becomes important. However, it is evident that in the resolved photoproduction
  the relation between
 the fusion and fragmentation mechanisms will be the same as in the direct photoproduction.

\section{Conclusions}
\label{sec:five}

It was shown that working within NRQCD both in the collinear
parton model and the QMRK approach, we have predicted the dominant
role of the fusion mechanism as compared to the
fragmentation one in the $J/\psi$ meson photoproduction at all
relevant values of charmonium transverse momenta. Therefore, it
is impossible to extract the $c-$quark or gluon fragmentation function
into $J/\psi$ meson from the photoproduction data similar as in
the hadroproduction  processes, as it was shown previosly
\cite{PRD2003}. Opposite to the assumption in the collinear
factorization,  that the fragmentation contribution is enhanced by
powers of $(p_T^2/m_c^2)$ relative to the  contribution of the fusion
model, we have demonstrated that such an enhancement is absent  in the
non-collinear factorization approach, QMRK. We have obtained
equal slopes of the $p_T-$spectra for both production
mechanisms at the asymptotically large $p_T$ in the QMRK approach.

\section{Acknowledgements}

We thank B.~A.~Kniehl, G.~Kramer and D.~V.~Vasin for useful discussions.
V.~S. thanks the 2nd Institute for Theoretical Physics at the
University of Hamburg for the hospitality extended to him during his visit when
this research was carried out. The work of V.~S. was supported in part by a DAAD Grant  {A/06/27387}.
A.~S. is grateful to the International Center of Fundamental Physics in
Moscow and the Dynastiya Foundation for financial support.

\newpage

\newpage

\newpage

\newpage

\newpage

\newpage

\newpage

\end{document}